\documentclass[fleqn,usenatbib]{mnras}
\usepackage[T1]{fontenc}
\usepackage{ae,aecompl}
\usepackage{color}
\usepackage{graphicx,latexsym,amssymb,geometry,pdflscape,pdfpages,amsmath}
\usepackage{natbib}
\usepackage{lineno}
\newcommand\arcs{\ensuremath{^{\prime\prime}}}

\newcommand\arcm{\ensuremath{^\prime}}

\newcommand\src{SAX\,J1808.4--3658}
\newcommand{\fermi}{{\it Fermi}}

\title[SAX\,J1808.4$-$3658 in \emph{{\fermi}}-LAT]{{SAX\,J1808.4$-$3658,} an accreting millisecond pulsar shining in gamma rays?}

\author[E. de O\~na Wilhelmi et al.]{E.~de O\~na Wilhelmi,$^{1}$
  A.~Papitto,$^{1}$ J.~Li,$^{1}$
  N.~Rea,$^{1,2}$ D.~F.~Torres,$^{1,3}$
  \newauthor L.~Burderi,$^{4}$ T.~Di Salvo$^{5}$
  R.~Iaria,$^{5}$ A.~Riggio,$^{4}$ and
  A. Sanna$^{4}$
\\
$^{1}$Institute for Space Sciences (CSIC$-$IEEC),
              Campus UAB,  Carrer de Can Magrans s/n,
              08193 Barcelona, Spain\\
$^{2}$Anton Pannekoek Institute for Astronomy, University of Amsterdam,
Postbus 94249, NL-1090 GE Amsterdam, the Netherlands\\
$^{3}$Instituci\'o Catalana de Recerca i Estudis Avan\c{c}ats (ICREA),  08010 Barcelona, Spain\\
$^{4}$Universit\`a degli Studi di Cagliari, Dipartimento di Fisica,
  SP Monserrato-Sestu, KM 0.7, 09042 Monserrato, Italy\\
$^{5}$Dipartimento di Fisica e Chimica, Universit\`a di Palermo, via Archirafi 36, 90123 Palermo, Italy
}

\date{Accepted XXX. Received YYY; in original form ZZZ}
\pubyear{2015}

\begin{document}
\label{firstpage}
\pagerange{\pageref{firstpage}--\pageref{lastpage}}
\maketitle



\begin{abstract}
We report the detection of a possible gamma-ray counterpart of the
accreting millisecond pulsar \src. The analysis of $\sim$6 years
of data from the Large Area Telescope on board the \fermi\ Gamma-ray
Space Telescope (\fermi-LAT) within a region of 15\degr\ radius around the
position of the pulsar reveals a point gamma-ray source detected at
a significance of $\sim$6$\sigma$ (Test Statistic ${\rm TS} =  32$), with position
compatible with that of \src\ within 95\% Confidence Level. The
energy flux in the energy range between 0.6 GeV and 10 GeV amounts to
($2.1\pm0.5)\times10^{-12}$ erg cm$^{-2}$ s$^{-1}$ and the spectrum
is well-represented by a power-law function with photon index
$2.1\pm0.1$. We searched for significant variation of the flux at the spin frequency of the pulsar and for orbital modulation,
taking into account the trials due to the uncertainties in
the position, the orbital motion of the pulsar and the
intrinsic evolution of the pulsar spin. No significant deviation from
a constant flux at any time scale was found, preventing a firm
identification via time variability. Nonetheless, the association of
the LAT source as the gamma-ray counterpart of \src\ would match the
emission expected from the millisecond pulsar, if it switches on as a
rotation-powered source during X-ray quiescence.

\end{abstract}



\section{Introduction}

Accretion-powered millisecond pulsars (AMSPs) are neutron stars (NSs)
that orbit a low-mass companion star ($\la$ 1\,M$_{\odot}$) and show
coherent X-ray pulsations at periods of a few milliseconds during X-ray
flares known as outbursts, caused by the impact of an accretion stream
onto the NS surface. The coherent pulsations observed in the X-ray
light curve during outbursts are due to the channelling by the NS
magnetosphere of (at least part of) the accretion flow to the magnetic
poles of the NS.

  \src\ was the first AMSP discovered \citep{1998Natur.394..344W};
  since 1996, it has gone into a few-weeks long outburst eight times,
  i.e roughly every 2.5 years ({\citealt{2012arXiv1206.2727P} and
    references therein, and \citealt{sanna2015} for the most recent
    one) reaching a peak X-ray luminosity of a few times 10$^{36}$~erg~s$^{-1}$. During X-ray quiescence it shows a much fainter unpulsed
    X-ray emission, which attains a 0.5--10 keV luminosity ranging
    between 0.5 and $1\times10^{32}$ erg s$^{-1}$
    \citep{2002ApJ...575L..15C,2007ApJ...660.1424H}.

Millisecond pulsars (MSPs) are believed to achieve their fast rotation
during a Gyr-long phase of accretion of mass and
angular momentum from a companion star
\citep{1982Natur.300..728A,1982CSci...51.1096R}. When the
mass-transfer rate declines at the end of the accretion phase the NS
magnetosphere is able to expand up to the light cylinder, and the
pulsar switches on as an MSP powered by the rotation of its magnetic
field. MSPs accelerate electron/positron pairs along field lines,
driving a pulsed emission observed mainly in the radio and
gamma-ray bands. The close link between accreting NSs and MSPs was
recently demonstrated by the discovery of IGR
  J18245-2452, which during outburst was observed as an AMSP, after having been previously detected as a radio MSP during
  X-ray quiescence \citep{2013Natur.501..517P}. This source switches
between these two states over a few weeks, presumably in response to
variations of the mass in-flow rate
\citep{1994ApJ...423L..47S,2002ApJ...575L..15C,2003A&A...404L..43B}. Several
indirect indications have been collected that a radio pulsar turns on
during the quiescent state of other AMSPs. For \src, in
particular, the amount of optical light reprocessed by the companion
star during X-ray quiescence \citep{2001MNRAS.325.1471H} is compatible
with irradiation by a radio pulsar \citep{2009A&A...496L..17B}, the
decrease of the NS spin period between consecutive outbursts is
similar to the rate observed from MSPs
\citep{2009ApJ...702.1673H,2012ApJ...746L..27P}, and the rapid
increase of the orbital period suggests ejection of the mass
transferred from the companion star and/or changes in the mass
quadrupole moment of the companion
\citep{2008MNRAS.389.1851D,2009A&A...496L..17B,2012ApJ...746L..27P}. However,
radio pulsations have not been detected from either \src\ or other
AMSPs
\citep{2003ApJ...589..902B,2009A&A...497..445I,2010A&A...519A..13I},
except for the case of IGR J18245-2453
\citep{2013Natur.501..517P}. This could be due to an intrinsic low
luminosity of the radio pulsar, geometrical effects, and/or free-free
absorption from material ejected from the system by the pulsar
radiation pressure \citep{2008MNRAS.389.1851D}. Note that two more
MSPs, PSR J1023+0038 \citep{2009Sci...324.1411A,2014ApJ...781L...3P}
and XSS J12270-4859 \citep{2014MNRAS.441.1825B}, have been observed in
an intermediate state characterized by the presence of an outer
accretion disk, and during which accretion-powered X-ray pulsations
were detected \citep{2014arXiv1412.1306A,2015MNRAS.449L..26P}. However
the X-ray luminosity of these sources during such episodes
($\approx5\times10^{33}$ erg s$^{-1}$) is much fainter than that
usually attained by AMSPs, which possibly indicates that a large
fraction of the mass in-flow is ejected by the quickly rotating NS
magnetosphere rather than accreted onto the NS surface.

Turning to high energies (HE; 100 MeV $<$ E $<$ 100 GeV) is a
promising strategy to detect the emission expected from an MSP turned
on during the X-ray quiescent state of an AMSP. \emph{{\fermi}}-LAT
\citep{2009ApJ...697.1071A} has proved to be an efficient
rotation-powered MSP detector \citep{2013ApJS..208...17A} benefiting
from the larger emission angle in gamma rays, and the absence of absorption from material possibly
enshrouding the binary. The gamma-ray pulsar sample comprises not only
canonical young pulsars but also recycled MSPs, that generally show a similar spectral shape as the young ones. The sky region of
several AMSPs was already investigated by \citet{2013ApJ...769..119X}
to search for gamma-ray emission in the 100 MeV to 300 GeV range over
a time span of 4 years, but they did not detect significant emission
associated with any AMSP. A source compatible with the position of
\src\ and dubbed 3FGL J1808.4$-$3703, is listed in the recently
published \emph{{\fermi}}-LAT 4-year point source catalogue
(3FGL, \citealt{2015arXiv150102003T}) with a detection significance of
4.5$\sigma$. Also, a possible detection of the gamma-ray
counterpart of \src\ was reported by \citet{2015arXiv150200733X}, who
nevertheless searched for gamma-ray pulsations using only the nominal
values of the ephemeris reported in \citealt{2009ApJ...702.1673H},
apparently overlooking the effect of the uncertainties on the
position, orbital and spin parameters over the coherence of a signal
searched in a few years-long time series.  Here we analyze almost six
years of LAT data from the region around \src\ to
investigate the possibility of the source emitting a significant
fraction of its energy in the gamma-ray regime. We performed a
detailed timing analysis to search for periodic features that could
firmly identify the gamma-ray source as the counterpart of \src,
carefully treating the impact of the uncertainties of the system
timing and spatial parameters on the range of parameters that have to
be considered.

\section{Data analysis and results}

\subsection{Data Analysis}

To search for a gamma-ray counterpart of \src\ we analyzed data
obtained with \emph{{\fermi}}-LAT in a region of 15\degr\ radius around its
position (RA$_{\rm J2000}= 18\rm h 08\rm m 27.62\rm s$, ${\rm Dec}_{\rm
  J2000}=-36\degr\ 58\arcm 43.3\arcs$,
\citealt{2008ApJ...675.1468H}).
The LAT experiment on board the \emph{{\fermi} Gamma-ray Space
  Telescope} satellite is sensitive to gamma rays with energies from
20 MeV to above 300 GeV, recording events with a timing accuracy better than 1 $\mu$s
\citep{2009APh....32..193A}. Almost six years of data (\emph{P7REP}, SOURCE class) obtained between the beginning of the
operation MJD\,54682.6 (August 4, 2008) and MJD\,56812.4 (June 4,
2014) were processed using the publicly available \emph{{\fermi}}-LAT
Science Tools (software version v9r32p5), analyzed with the response
functions \emph{P7REP$\_$SOURCE$\_$V15} and using the templates
for Galactic (gll$\_$iem$\_$v05.fits) and isotropic
(iso$\_$source$\_$v05.txt) backgrounds. We selected data in the
  100 MeV to 300 GeV energy range. Standard time event cuts performed with the tool gtmktime (DATA$\_$QUAL==1, LAT$\_$CONFIG==1
    and ABS(ROCK$\_$ANGLE)$<$52) were applied. To suppress the effect
  of the Earth limb background, we excluded
time intervals when the Earth was in the field of view 
(FoV, when the LAT Z-axis was more than 52\degr\ from zenith),
and those in which part of the selected ROI was observed with zenith
angle larger than 100\degr.
A second analysis was performed
excluding the periods in which \src\ was in an X-ray bright outburst
state, i.e. between September 22nd and November 7th 2008 and November
5th to 20th 2011 \citep{2009ApJ...702.1673H,2012ApJ...746L..27P}, with
compatible results to those presented next. No significant
  gamma-ray excess is detected when analysing time intervals when the
  source was in outburst.
  
\begin{figure}
\centering
\includegraphics[width=\linewidth, angle=0]{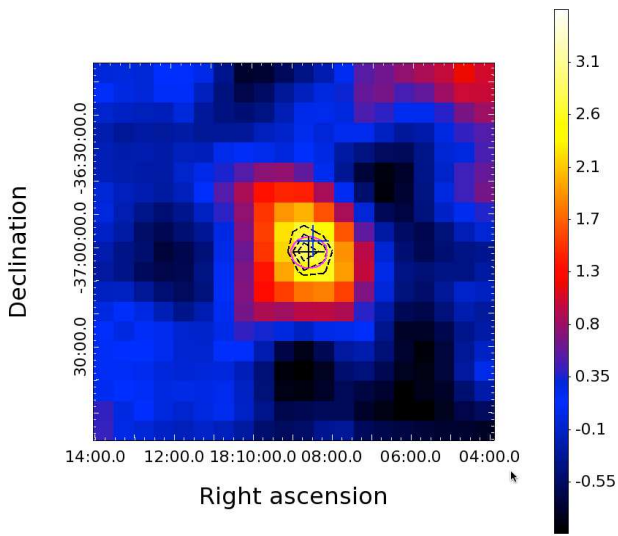}
\caption{ {\fermi}-LAT residual 2\degr$\times$2\degr\ (using a pixel size of $0.1\degr\times0.1\degr$) count map above 1 GeV of the
  SAX J1808.4--3658 region smoothed with a Gaussian of width
  $\sigma=0.3\degr$ (units of the scale on the right are
  counts). The best-fit position of the gamma-ray source is marked
  with a black cross whereas the position of SAX J1808.4--3658 is
  marked in blue. The black-dashed lines show the TS significance
  contours above 1 GeV corresponding to CL of 68\%, 95\% and 99\%. The magenta circle shows the 95\% CL error in the best-fit position.}
\label{fig1}
\end{figure}

\begin{figure}
\centering
\includegraphics[width=\linewidth, angle=0]{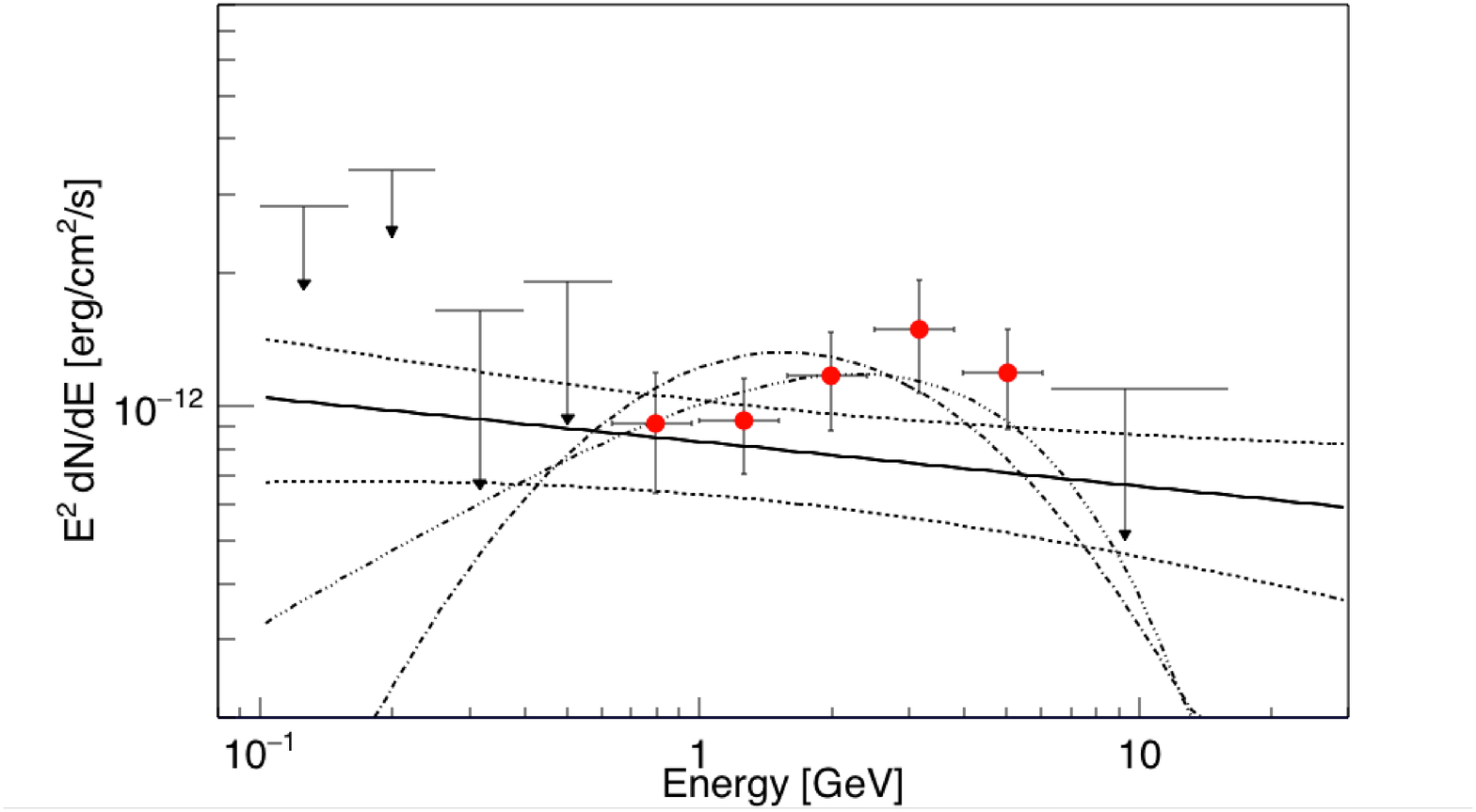}
\caption{SED obtained at the position of SAX J1808.4--3658. The best-fit power law function is plotted as a
  solid line, and the dashed lines shows the statistical errors in
  the global fit. No highly significant improvement (by less than 3$\sigma$) is obtained using more
  sophisticated models, such as a power law with cut-off at high energies function or a log parabola function. The best-fit functions for the latter models are shown in dash-dot-dot and dash-dot lines, for exponential cutoff and log parabola functions respectively. }
\label{fig2}
\end{figure}

\subsection{Image and Spectral analysis}
The image of the sky around the position of \src\ was obtained in the
energy range above 1 GeV, where the angular resolution of
\emph{{\fermi}}-LAT reaches
$\approx$0.8\degr\footnote{http://www.slac.stanford.edu/exp/glast/groups/canda\\/lat$\_$Performance.htm} on axis. Using
the same exposure and instrument response from the data set, we modelled the region of interest (ROI) with all the 3FGL sources \citep{2015arXiv150102003T} (excluding 3FGL J1808.4$-$3703) and the standard Galactic
and isotropic diffuse components and we let all the spectral parameters free (except for the spectra of the sources more than 10\degr\ away from the center) in a maximum likelihood fit (using \emph{gtlike}\footnote{http://fermi.gsfc.nasa.gov/ssc/data/analysis/scitools/}). Figure \ref{fig1} shows the residuals
with respect to the best-fit model. The residual image shows a point source compatible
with the position of \src\ (see blue cross). For the computation of
the significance and spectral parameters of \src\ we added a
point source at ${\rm RA}_{\rm J2000}=272.115\degr$ and ${\rm Dec}_{\rm
  J2000}=-36.98\degr$ to account for the gamma-ray excess. 

The inclusion of a point gamma-ray source at the position of
\src\, described with a power-law function,
$\phi_{\rm0}\times$(E/E${\rm_0}$)$^{-\Gamma}$, normalised at ${\rm E}_{\rm
  0}=1.44\,{\rm GeV}$ with $\phi_{\rm0}=(2.42\pm0.56)\times10^{-10}\,{\rm GeV}^{-1}{\rm cm}^{-2}{\rm s}^{-1}$ and $\Gamma=2.1\pm0.1$ (in black in
Fig. \ref{fig2}), results in a test statistic (TS,
\citealt{1996ApJ...461..396M}) value of 31.6, which corresponds to a
source detection at a confidence level of $\sim6\sigma$. Applying
the \emph{pointlike maximum-likelihood fitting} package
\citep{2010PhDT.......147K} we fit the position of the gamma-ray
excess above 100 MeV to ${\rm RA}_{\rm J2000}=(272.143\pm0.037)\degr$ and ${\rm Dec}_{\rm
  J2000}=(-37.034\pm0.032)\degr$ (compatible with the position of
\src\ within 95\% CL, see magenta circle in Fig. \ref{fig1}). The contour
lines for 68, 95 and 99\% Confidence Level (CL) obtained from the significance map,
calculated with \emph{gttsmap} for events above 1~GeV are also shown.

The spectral energy distribution (SED) for a point source
centered on the position of \src\ was derived by means of a binned
likelihood fitting, divided in 15 logarithmic bins between 100 MeV and
300 GeV (see Fig. \ref{fig2}). The spectral points obtained for each
energy bin (with significance of more than 2$\sigma$), fitting the data with a power-law
function with fixed photon index of 2, are shown in
Fig. \ref{fig2}. More sophisticated spectral shapes, aiming to fit the 100 MeV to 300 GeV spectral range, do not provide a
statistically-significant improvement to the fit. The comparison with
a fit to a log parabola function ($\phi=\phi_{\rm0}$(E/E$_{\rm
  b}$)$^{-(\alpha+\beta log(\rm E/E_{\rm b}))}$) results in a
difference in the maximum likelihood of 2$\times$$\Delta{\rm L}/\Delta({\rm ndf}) = 10$ (with $\Delta({\rm ndf})=2$ and probability ${\rm P}=6.7\times10^{-3}$, corresponding to 2.7$\sigma$),
whereas the comparison to a power-law function plus exponential cutoff
($\phi=\phi_{\rm0}$(E/E$_{\rm 0}$)$^{-\gamma_{1}}$exp$(-(E/E_{\rm
  c})$)) leads to an increase of only 8 (with $\Delta({\rm ndf})=1$ and probability ${\rm P}=4.57\times10^{-3}$, corresponding to 2.8$\sigma$). For the log parabola hypothesis, the best-fit parameters we found are $\phi_{\rm0}= (4.20\pm0.32)\times10^{-10}\,{\rm GeV}^{-1}{\rm cm}^{-2}{\rm s}^{-1}$, $\alpha=1.91\pm0.09$ and $\beta=0.41\pm0.02$ for a break energy of $E_{\rm b}=1.4$ GeV, whereas when we try to fit to a power-law function plus exponential cutoff we obtain $\phi_{\rm0}= (3.2\pm2.7)\times10^{-8}\,{\rm GeV}^{-1}{\rm cm}^{-2}{\rm s}^{-1}$, $\gamma=1.4\pm0.4$ and $E_{\rm c}=3.8\pm25$ GeV.

\begin{table*}
\caption{Parameters used in the periodicity search.}
\label{table:1}

\begin{tabular}{lcccc}
\hline
Parameter & Best estimate, $x_i$ & Error, $\sigma_{x_i}$ & Sensitivity, $\delta_{x_i}$ &No. corrections, $N_{x_i}$ \\
\hline
Ecliptic longitude, $\lambda$ & $271.737918\degr$ & $0.13\arcs$ & $0.015\arcs$ & 16 \\
Ecliptic latitude, $\beta$ & $-13.552162\degr$ & $0.15\arcs$ & $0.064\arcs$ & 5 \\
Orbital period, $P_{orb}(T_0)$ & $7249.157964$ s & $7.6\times10^{-5}$ s & $3.4\times10^{-4}$ s &  1\\
Orbital period derivative, $\dot{P}_{orb}(T_0)$ & $3.17\times10^{-12}$ & $0.70\times10^{-12}$ & $1.8\times10^{-12}$ & 1 \\
Orbital period second deriv., $\ddot{P}_{orb}(T_0)$ & $1.65\times10^{-20}$ s$^{-1}$& $0.35\times10^{-20}$ s$^{-1}$& $2.0\times10^{-20}$ s$^{-1}$& 1 \\
Epoch of passage at asc. node, $T^*$ & $54729.999079$ MJD & $0.78$ s & $14.9$ s & 1\\
Semi-major projected axis, $a\,\sin{i}/c$ & $62.812$ lt-ms & $2\times10^{-3}$ lt-ms & $2.0$ lt-ms & 1\\ 
Spin frequency, $\nu(T_0)$ & $400.97521014$ Hz & $2.4\times10^{-2}\,\mu$Hz & $5.4\times10^{-3}\,\mu$Hz &  27 \\
Spin frequency derivative, $\dot{\nu}$ & $7.1\times10^{-16}$ Hz/s & $-1.2\times10^{-16}$ Hz/s & $3.3\times10^{-17}$ Hz/s  & 25 \\
 \hline
\multicolumn{5}{p{\textwidth}}{Notes: Position and orbital parameters were
  taken from \citet{2008ApJ...675.1468H,2009ApJ...702.1673H} and
  \citet{2012ApJ...746L..27P}. Frequency and frequency derivative were
  measured by fitting the spin frequency values given by
  \citet{2008ApJ...675.1468H,2009ApJ...702.1673H,2012ApJ...746L..27P}
  with a constant spin-down function (see black dashed line in
  Fig.~\ref{fig3}). The reference epoch is
  $T_0=\mbox{MJD}\,54730$. The total observation time is
  $T_{obs}=2129.8$ d. See Sec. 2.2 for details on the assessment of the sensitivity to a coherent signal over the considered time series.}

 \end{tabular}
\end{table*}

\subsection{Timing analysis}

The long term exposure-corrected (counts/exposure) light curve, produced by means of \emph{aperture
  photometry} within a radius of 1\degr and retaining photons
between 100\,MeV and 300\,GeV, does not show any deviation from a
constant flux. The fit to a constant flux results in a $\chi^2$/ndf of 0.9 (for ${\rm ndf}=210$) using a linear binning with a bin width of 10 days.
 In order to search for a modulation of the gamma-ray flux at the orbital
period of the gamma-ray flux, we folded the light curve in 10 bins
around the value of the orbital period predicted according to:
\begin{equation}
\label{eq:orbit}
P_{orb}(t)=P_{orb}(T_0)+\dot{P}_{orb}(T_0)\times(t-T_0)+\frac{1}{2}\ddot{P}_{orb}(T_0)\times(t-T_0)^2,
\end{equation}
where $T_0=54730$ MJD, and the values of $P_{orb}(T_0)$,
$\dot{P}_{orb}(T_0)$ and $\ddot{P}_{orb}(T_0)$ are listed in Table 1 \citep{2008MNRAS.389.1851D,2009A&A...496L..17B,2012ApJ...746L..27P}.
The variance of the folded profile extracted considering photons at
energies larger than 100 MeV is $\chi^2/{\rm ndf}=11/9$, indicating no
evidence of a significant modulation. A similar result is obtained
considering only photons at higher energies (e.g., $>2$~GeV). The fit to a constant function results in a $\chi^2$/ndf of 30/9 (see Fig. \ref{orb}), corresponding to a marginal P-value of $5\times10^{-4}$ pre-trial.

\begin{figure}
\centering
\includegraphics[width=0.8\linewidth, angle=0]{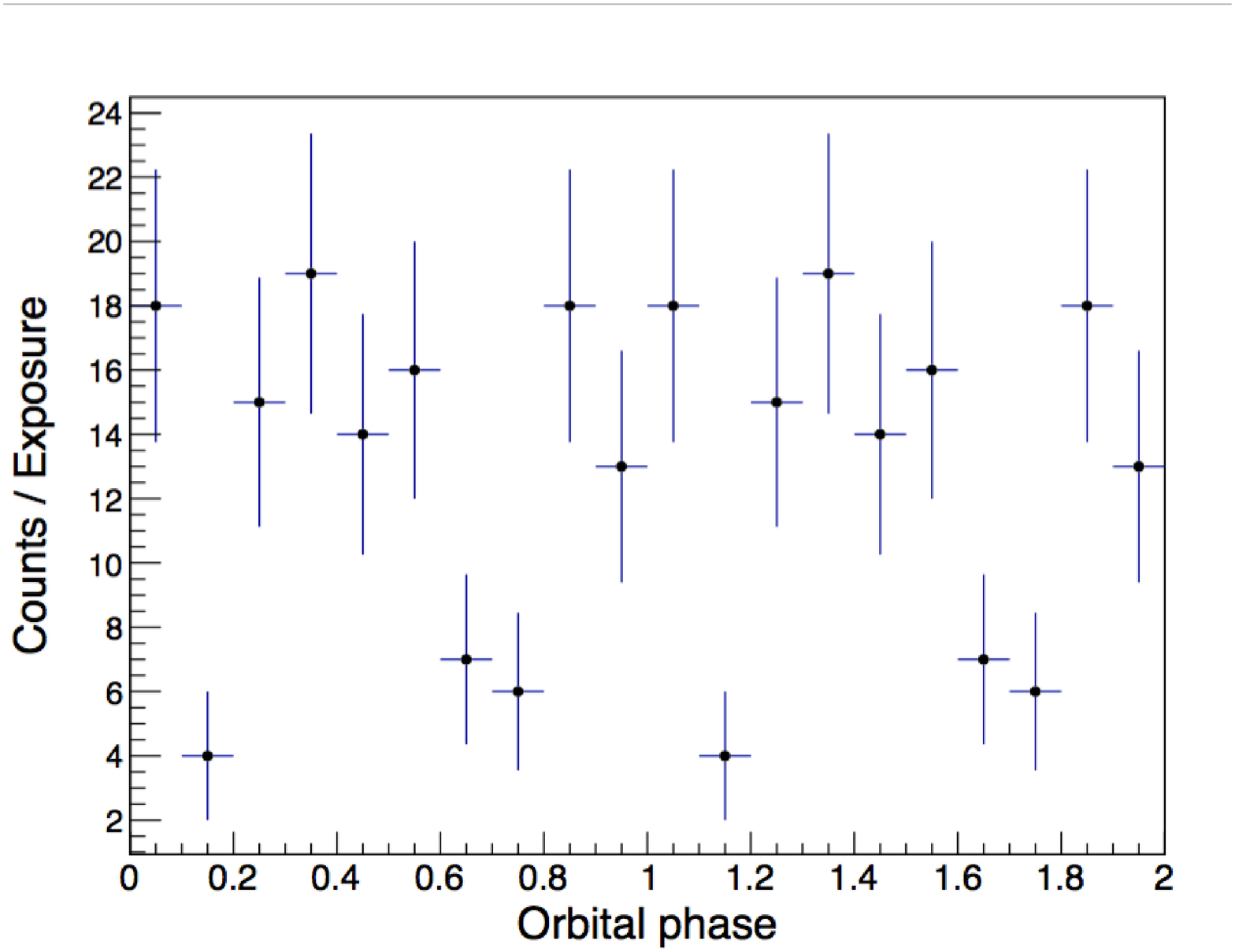}

\caption{Phaseogram obtained folding the arrival time of the exposure-corrected gamma-ray
  photons ${\rm E}>2\,{\rm GeV}$ with the orbital period P$_{\rm orb}$(t) obtained from equation \ref{eq:orbit}.}
\label{orb}

\end{figure}

The search for gamma-ray pulsations is limited by the uncertainties in
the ephemeris and position of the source and the (faint) flux level
detected. This search must take into account the loss of signal
coherency over the $T_{obs}=2129.8$ d-long interval spanned by the observations. The following effects are considered, (i) the
orbital motion of the pulsar, (ii) the intrinsic evolution of the
pulsar spin, and (iii) uncertainties on the position of the source.

The frequency variations induced by the orbital motion of the source
have to be taken into account by correcting the photon arrival times to the line of
nodes of the binary system, using the most updated orbital solution
available (see Table~\ref{table:1}).
Assuming that the orbital period evolution is described by Eq.~\ref{eq:orbit}, the maximum uncertainty on the estimate of the orbital period driven by the errors on the values of $P_{orb}(T_0)$, $\dot{P}_{orb}(T_0)$ and $\ddot{P}_{orb}$, is evaluated by standard error propagation over the length of the considered time series:
\begin{equation}
\sigma_{P_{orb}}^{max}=\left[\sigma_{P_{orb}}^2+(\sigma_{\dot{P}_{orb}}T_{obs})^2+\left(\frac{1}{2}\sigma_{\ddot{P}_{orb}}{T_{obs}}^2\right)^2\right]^{1/2}=1.6\times10^{-4}\,\mbox{s}.
\end{equation}
 In order to check if the actual uncertainties on the orbital
 parameters produce a loss of coherence of the pulsar signal, we used
 the expressions given by \citet[][see Table~3 therein]{2012MNRAS.427.2251C}, who estimated the fraction of power
 lost, $\epsilon$, as a function of the difference between the actual
 value of an orbital parameter  and the one used to refer photon
 times of arrival to the line of nodes of the binary system. We
 evaluated the analytical relations they give for the case of a
 circular orbit:
\begin{eqnarray}
\delta{(a\sin{i}/c)}=&\frac{1}{2\nu_0}\frac{1}{\epsilon^2}\\
 \delta{T^*}=&\frac{0.1025  P_{orb}}{\pi\nu_0 (a\sin{i}/c)}\frac{1}{\epsilon^2}\\ 
\delta{P_{orb}}=&\frac{{P_{orb}}^2}{2\pi\nu_0
  (a\sin{i}/c) T_{obs}} \sqrt{\left(\frac{1-\epsilon^2}{10}\right)},
\end{eqnarray}
for $\epsilon=0.8$, obtaining the values listed in Table~\ref{table:1}
in the column labelled as `Sensitivity'. Here, $a\sin{i}/c$ is the
projected semi-major axis of the NS orbit, and $T^*$ is the epoch of
passage of the NS at the ascending node of the orbit. For each of the
orbital parameters, $x_i$, the sensitivity value $\delta_{x_i}$ is
larger than the uncertainty $\sigma_{x_i}$ (see Table~\ref{table:1}),
assuring that signal coherence is not lost throughout the length of
the considered observation. The sensitivity to the uncertainty on the value of the first and second derivative of the orbital period were evaluated by taking the value that alone would produce a period shift equal to $\delta P_{orb}$, namely, $\delta \dot{P}_{orb}=\delta P_{orb}/T_{obs}$ and $\delta \ddot{P}_{orb}=2 \delta P_{orb}/T_{obs}^2$ (see Eq.~1).

So far, coherent pulsations were detected from {\src} only in the
X-ray light curves observed during six of the outbursts shown by the
source since 1998, each of which lasted a few weeks. We 
extrapolated the spin evolution of the pulsar, fitting a constant
spin-down model to the frequencies measured in different outbursts.
In addition, the measure of the spin frequency of {\src} during
each of the outbursts is complicated by the presence of strong timing
noise that exceeds Poisson counting noise and affects, to a different
extent, the first and second harmonic of the signal
\citep{2009A&A...496L..17B}. \cite{2009ApJ...702.1673H} and
\cite{2012ApJ...746L..27P} measured the frequency of the signal during
each of the outbursts using a frequency-domain filter to weigh the
harmonics according to the observed noise properties and estimated
the uncertainty by performing Monte Carlo simulations. Using this method
they found no significant evolution of the spin frequency during the
various outbursts. Here we consider the spin frequency that they
measured in each of the outbursts, summing in quadrature the
uncertainty driven by positional errors $\delta\nu_{pos}^{max}$ (see
below) to the uncertainty quoted on their values. By fitting the
average frequency values observed during the six different outbursts
with a constant spin-down trend,
\begin{equation}
\nu(t)=\nu(T_0)+\dot{\nu}\times(t-T_0),
\end{equation}
 we estimated the spin frequency derivative as
 $\dot{\nu}=(7.1\pm1.2)\times10^{-16}$ Hz s$^{-1}$ (see dashed line in
 Fig.~\ref{fig3}), compatible with the value given by
 \cite{2012ApJ...746L..27P}. Propagating the errors on the spin
 frequency and its derivative over the whole length of the
 observations leads to a maximum uncertainty on the signal frequency
 of 
\begin{equation}
\sigma_{\nu}^{max}=[\sigma_{\nu}^2+(\sigma_{\dot{\nu}}T_{obs})^2]^{1/2}\simeq 3\times10^{-2}\,\mu\mbox{Hz}.
\end{equation}
We assume that the minimum difference between frequencies that
produces a significant power loss in a search for a signal is equal to
the independent Fourier frequency spacing,
$\Delta\nu_{IFS}=1/T_{obs}=5.4\times10^{-3}\,\mu\mbox{Hz}$. As the
maximum uncertainty on the spin frequency $\sigma_{\nu}^{max}$ is
larger than $\Delta\nu_{IFS}$, we are forced to perform a search over
different possible values of $\nu(T_0)$ and $\dot{\nu}$ in order to
avoid a significant loss of signal power. We varied the spin frequency
and its derivative in steps equal to the amount that produces an
uncertainty equal to $\Delta\nu_{IFS}$ over a time interval equal to
$T_{obs}$,
i.e. $\delta\nu=\Delta\nu_{IFS}=5.4\times10^{-3}\,\mu\mbox{Hz}$, and
$\delta{\dot{\nu}}=\Delta\nu_{IFS}/T_{obs}=1/T_{obs}^2=2.9\times10^{-17}$ Hz/s. In
order to cover an interval equal to $\pm 3\sigma$ around the central
value of $\nu(T_0)$ and $\dot{\nu}$, we then performed
$N_{\nu}=2\times(3\sigma_{\nu})/\delta\nu=27$ and
$N_{\dot{\nu}}=2\times(3\sigma_{\dot{\nu}})/\delta\dot{\nu}=25$
correction trials on the spin frequency and its derivative,
respectively.  The limits of the range of values covered is plotted in
Fig.~\ref{fig3} using blue dashed lines.

\begin{figure}
\centering
\includegraphics[width=0.8\linewidth, angle=0]{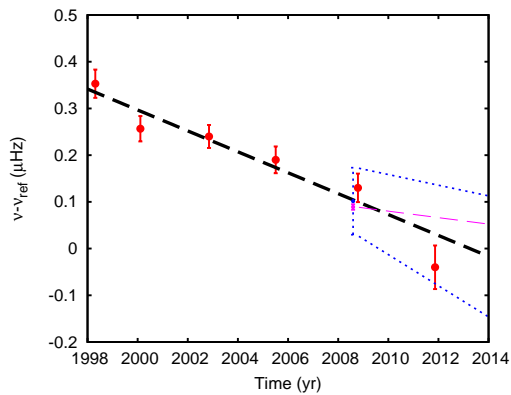}

\caption{Evolution of the spin frequency of SAX J1808.4$-$3658 as observed in the
  X-ray band. The dashed black line is the best-fitting spin-down
  trend, the dotted blue lines mark the range of parameters searched,
  and the magenta line is the solution that gives the maximum
  H-test value. The reference frequency is $\nu_{o}=400.975210$ Hz.}
\label{fig3}

\end{figure}

\begin{figure*}
\centering
\includegraphics[width=0.5\linewidth, angle=0]{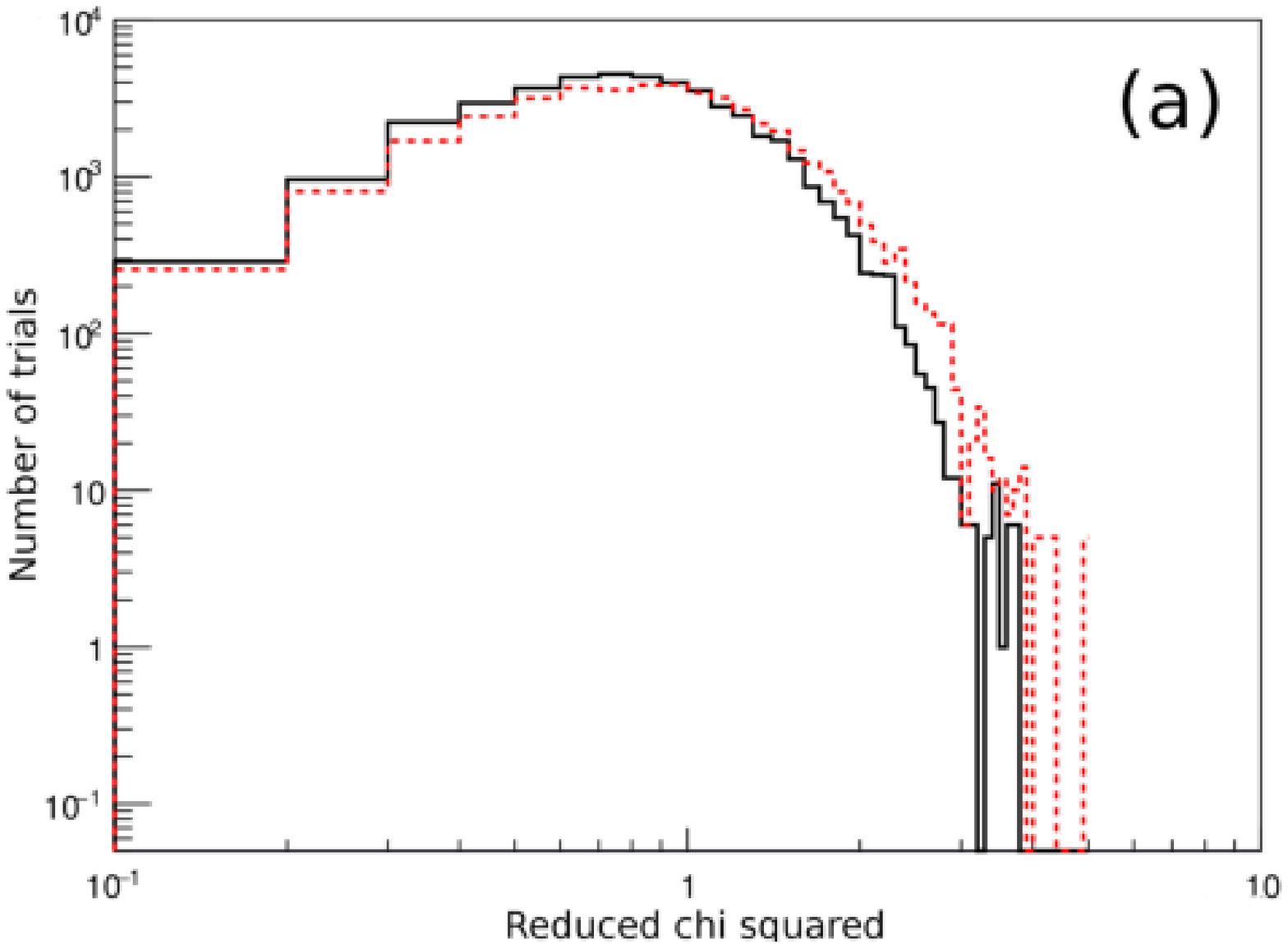}
\includegraphics[width=0.48\linewidth, angle=0]{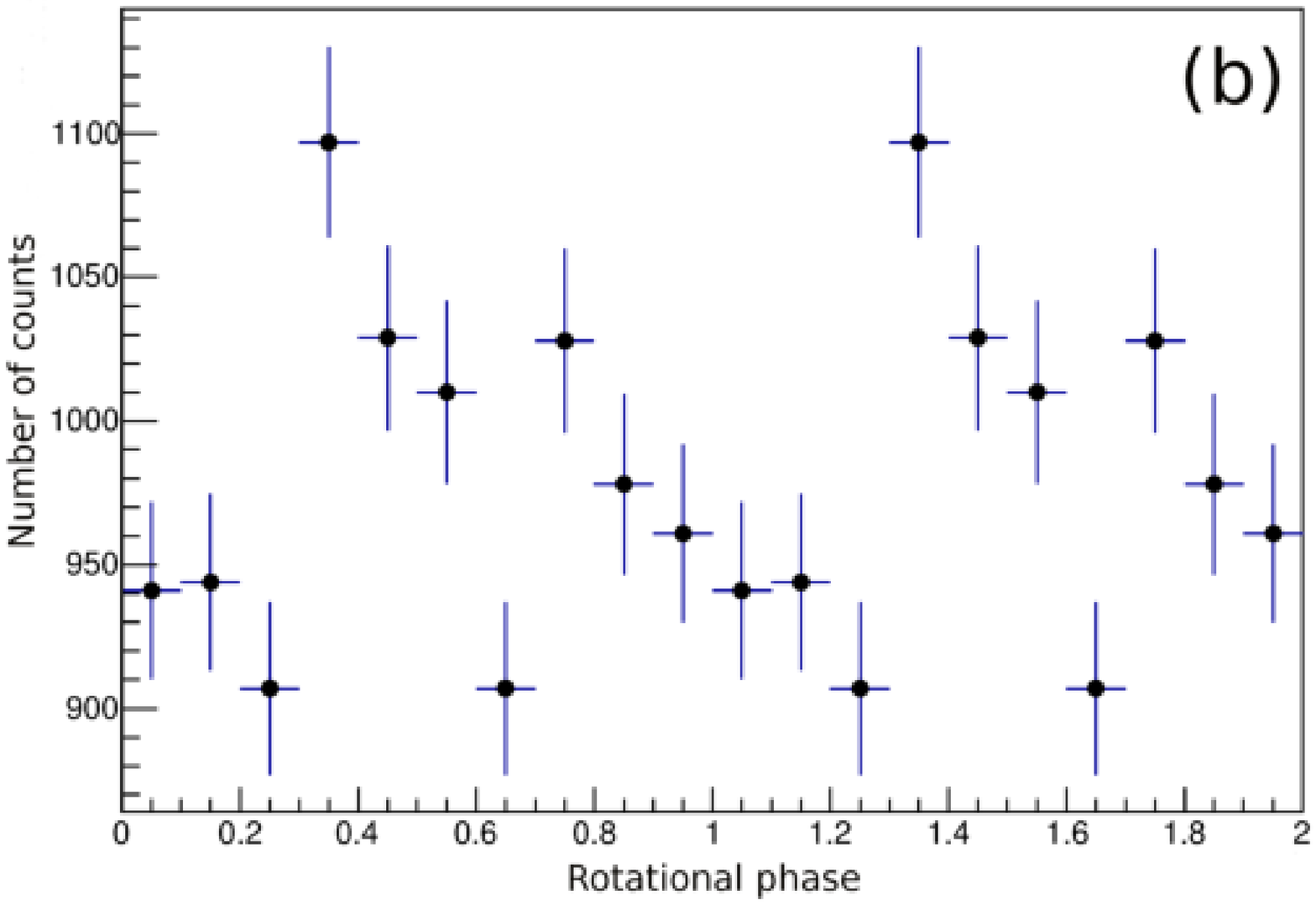}

\caption{Left a) $\chi^2$/ndf distribution for positions, independent frequencies and frequency derivatives tested
  in a range of 3$\sigma$ around the expected $\nu$ and $\dot{\nu}$
  and of 1$\sigma$ around the central estimate of the coordinates. The
  black distribution shows the results when using the correct ephemeris
  whereas the red one refers to the \emph{fake} ones. Right b)
  Phaseogram obtained folding the arrival time of the gamma-ray
  photons with $\nu_{\rm 0}$ and $\dot{\nu}_{\rm 0}$ and on the
  position $\lambda_{\rm 0}$ and $\beta_{\rm 0}$, which results in a
  maximum in the two periodicity tests applied. Two cycles are plotted for
clarity, and y-axis is zero-supressed.}
\label{fig4}
\end{figure*}

An additional number of correction trials has to be performed when
photon arrival times are converted to the Solar System barycenter,
because of the uncertainty on the source position. We considered the
position of the optical counterpart determined by
\cite{2009ApJ...702.1673H}, RA = 18h08m27.62s,
DEC = -36\degr58\arcm43.3\arcs, with an uncertainty of 0.15\arcs\ 
[corresponding to ecliptic coordinates $\lambda=271.737918\degr$,
  $\beta=-13.552162\degr$, affected by uncertainties
  $\sigma_{\lambda}=013\arcs$, and $\sigma_{\beta}=0.15\arcs$,
  respectively (see Table~\ref{table:1})].  A difference of
($\delta\lambda$, $\delta\beta)$ between the actual ecliptic
coordinates of the source ($\lambda$, $\beta$), and those used to
correct the time series, yields an apparent modulation of the spin
frequency of the signal equal to:
\begin{equation}
\label{eq:pos}
\delta\nu_{pos}=\nu y \left(\frac{2\pi}{P_{\oplus}}\right)[\cos A_0\, cos
  \beta\, \delta\lambda + \sin A_0\, \sin\beta\, \delta\beta]
\end{equation}
\citep{1972ApJ...173..221M}.  Here $y$ is the Earth distance from the
Solar System barycenter, $P_{\oplus}=1$ yr,
$A_0=[2\pi(T_0-T_{\gamma})/P_{\oplus}]$, $T_0$ is the start time of
observations and $T_{\gamma}$ is the Vernal point. Considering
that our time-series covers $\simeq 5.8$ yr, the uncertainty on the
position of \src\ translates into a modulation of the signal frequency
of amplitude
$\delta\nu_{pos}^{max}\simeq2.4\times10^{-2}\,\mu\mbox{Hz}$. As for
the uncertainty on the spin frequency and its derivative (see above),
this value is also larger than the spacing between independent Fourier
frequencies, $\Delta\nu_{IFS}=5.4\times10^{-3}\,\mu\mbox{Hz}$, forcing
us to perform a series of corrections on the coordinates used to
barycenter the \emph{{\fermi}}-LAT light curve. We estimated the minimum
difference between coordinates that produces a significant signal
loss as that producing a frequency oscillation $\delta\nu_{pos}$
(evaluated using Eq.\ref{eq:pos} and putting $\cos A_0$ and $\sin A_0$
equal to one, for simplicity) equal to $\Delta\nu_{IFS}$. We thus
obtained $\delta\lambda=0.015\arcs$ and $\delta\beta=0.064\arcs$.  In
order to cover a range within 1$\sigma$ from the central estimates of
the source coordinates,
$N_{\lambda}=2\sigma_{\lambda}/\delta_{\lambda}=16$ and
$N_{\beta}=2\sigma_{\beta}/\delta_{\beta}=5$ preliminary corrections of
the time series were then performed (see Table~\ref{table:1}, where
the parameters of the grid used in the periodicity search are given).
That implies $N_{\lambda}N_{\beta}=80$ time series, for which $N_{\nu}$$N_{\dot{\nu}}=675$ searches over $\nu$ and ${\dot{\nu}}$ should be performed. Considering the
flux level of the detected source (with $\sim$100 photons detected
from the source direction), the
total number of trials needed to apply ($N_{tr}=54000$) strongly
hampers the search for gamma-ray pulsations at the spin period of the
source, given the current uncertainties and instrument
sensitivity. Considering these values, only a signal with a sinusoidal
amplitude $\ga 65\%$ (i.e. giving a $\chi^2/(\mbox{ndf}-1)=5.75$ for
$\mbox{ndf}=10$) would be detected at $3\sigma$ confidence level by an
epoch folding search technique, performed by sampling the profile with
${\rm ndf}=10$ phase bins \citep{1987A&A...180..275L}\footnote{We note that
  the sensitivity to a signal with a lower duty-cycle, like those
  often observed from radio pulsars would be higher than in the case of a
  sinusoidal signal.}

Nevertheless, we searched for a periodic signal in the \emph{{\fermi}}-LAT light
curve in the range considered in Fig. \ref{fig3}. The arrival time of
each event was first transformed to the Solar System barycenter using the
grid of positions determined previously, then we applied the
corrections for the orbital motion, and finally we calculated the
phase of each photon using the grid of values of frequency and
frequency derivative determined above. The time correction was done
using the LAT \emph{gtpphase} tool.  The uniformity of the phaseogram
is tested by both applying a simple epoch folding search test on a
10-bin pulse profile and an H-test \citep{2010A&A...517L...9D} on the
arrival events. Fig. \ref{fig4}a shows the distribution of the
$\chi^2$ obtained with an epoch folding search for different
positions, $\nu$ and $\dot{\nu}$. The two tests reached a maximum of
their value $\chi^{2}/{\rm ndf} =3.8$ (with $\mbox{ndf}=9$) and ${\rm H}=11.7$ for
${\rm m}=2$, where ${\rm m}$ is the number of harmonics used when the data set is
folded using the combination of $\lambda_{\rm 0}=271.737942\degr$,
    $\beta_{\rm 0}=-13.552197\degr$, $\nu_{\rm 0}=400.975210089$
Hz and $\dot{\nu_{\rm 0}}=-2.2\times10^{-16}$Hz/s.  We performed
the same statistics test around \emph{fake} values of $\nu$ and
$\dot{\nu}$ ($\nu_{\rm fake}=399.97521013$ Hz and $\dot{\nu}_{\rm
    fake}=-5.5\times10^{-16}$ Hz/s) to validate the uniformity of
the test (in red in Fig. \ref{fig4}a). \cite{2010A&A...517L...9D} showed that
the probability distribution for the H-test can be described by
${\rm P(}>{\rm H})={\rm exp(-0.4H)}$. From this expression we can derive a probability
of ${\rm P(}> 11.7\rm{)}=9.3\times10^{-3}$ before trials for the light curve to
deviate from a flat distribution. The folded light curve obtained with
$\nu_{\rm 0}$ and $\dot{\nu_{\rm 0}}$ and the position ($\lambda_{\rm
  o}$,$\beta_{\rm 0}$) is shown in Fig. \ref{fig4}b (two cycles are
plotted for clarity), and the relevant source spin evolution is
plotted as a magenta dashed line in Fig.~\ref{fig3}. Also considering
the large number of trials made (${\rm N}_{tr}=54000$) such a solution is
not significant. Similarly, the probability of obtaining a chi-squared
value of $\chi^{2}/{\rm ndf}=3.8$ for ${\rm ndf}=9$ in a single epoch-folding is
${\rm P}=8.2\times10^{-5}$. Considering all the trials made, we
expect ${\rm N}_{tr}\times P\simeq4.4$ folded profiles to yield such a
chi-squared value by chance, which indicates clearly that the detection
is not significant.


\section{Discussion}

The best-fit position of the gamma-ray source discovered is
located 3.2\arcm\ from the optical position of \src\ (within the 95\% CL of the gamma-ray source position). We investigated
a region of 0.15\degr\ radius surrounding the position of the gamma-ray source, and
no obvious possible gamma-ray-producing counterpart or gamma-ray accelerator was found beside the
AMSP. The only source detected in the surrounding is the radio galaxy
NVSS 180824$-$365813 \citep{1998AJ....115.1693C}, although the lack of
an X-ray counterpart and faint flux make it an unlikely
candidate to emit in gamma rays \citep{2014cosp...40E.239B}. A more detailed investigation of faint X-ray sources other than \src\ can be found in \citet{2015arXiv150200733X}.

In the 3FGL catalogue, a source compatible with the
position of \src\, and dubbed 3FGL J1808.4$-$3703 is listed. Its flux
and spectral parameters are compatible with the source reported
here. \citet{2015arXiv150200733X} reported similar investigations. Nevertheless a search for gamma-ray pulsations was done by the previous authors without taking into account the possible range of possible ephemeris. \citet{2015arXiv150200733X} also
reported a barely significant modulation at the orbital period. Here, a detailed timing analysis is performed, considering the uncertainties of the system timing and spacial parameters. We also checked that their result could be reproduced (with a statistical
significance in the 10 bins light-curve of $\chi^2/{\rm ndf}=30/9$,
corresponding to 3.5$\sigma$) only by extracting
photons with energies $>2$ GeV, coming from a region around
0.6\degr\ around the source. The non-detection of any significant
modulation when considering a region of different size or a different energy band raises doubts on the reliability of such a claim.

If the identification of the gamma-ray source found with the AMSP is
real, the gamma-ray emission could originate either in the pulsar
magnetosphere or in the intra-binary shock. No significant variation
at the time scale set by the orbital period has been found. We also
folded the arrival times of the gamma-ray photons around the spin
frequency of the pulsar, using the latest ephemeris measured during
the last flaring state, and allowing a deviation of 3$\sigma$ with
respect to the extrapolated value for $\nu$ and $\dot{\nu}$. The
position of the source was also varied to take into account the error
in the position determination when converting to the Solar System
barycenter. Considering the large number of trials needed to cover all
the possible spin and position parameters, no significant detection of
gamma-ray pulsations could be achieved.  Even though we can not yet
formally identify the LAT source with \src\, we can compute the
gamma-ray luminosity for a scenario in which \src\ is producing the
detected gamma-ray radiation at a distance of $3.5\pm0.1$~kpc
\citep{2006ApJ...652..559G}. We obtain a total luminosity of
L$_{\gamma}=(3\pm1)\times10^{33}$ erg/s in the energy range
between 0.6 GeV and 10 GeV (i.e, the energy range in which the source
was significantly detected, see Fig.~\ref{fig2}), which is compatible
with upper limits obtained previously by \citet{2013ApJ...769..119X}
in a search of gamma-ray counterpart of several AMSP, including
\src. If we compare with the total rotational power at present
($\dot{\rm E}=(1.1\pm0.2)\times10^{34}$ erg/s obtained from the values
of $\nu$ and $\dot{\nu}$ quoted in Table \ref{table:1} and using a
moment of inertia of 10$^{45}$g~cm$^{-2}$) assuming a beaming factor
of $f_\Omega=1$ \citep{2009ApJ...695.1289W}, we obtain an efficiency
of $\eta=L_{\gamma}/\dot{\rm E}\times100=(27\pm9)\%$, which is
within the range of efficiencies observed from MSPs detected at
high energy \citep{2013MNRAS.430..571E,2009arXiv0910.4707G,2013ApJ...763L..13R,2013ApJS..208...17A}. If
the association can finally be proven, SAX J1808.4$-$3658 in X-ray
quiescence will be similar to PSR\,J1311$-$3430
\citep{2012Sci...338.1314P,2013ApJ...763L..13R}, a fast MSP
($\approx2.5\,{\rm ms}$) in a compact binary system ($\approx2$\,h). The
spectral parameters are also compatible within the current statistics
to the ones found in other MSPs, with a hard spectrum and a turn over
at a few GeV \citep{2013MNRAS.430..571E,2013ApJS..208...17A}.

Two MSPs, PSR J1023+0038 and XSS J12270$-$4859, have recently been 
observed to switch between a rotation-powered radio pulsar state and
an intermediate state characterized by the presence of an outer
accretion disk. In the disk state these two sources showed a 0.1--100
GeV gamma-ray luminosity of a $\mbox{few}\times 10^{34}$ erg s$^{-1}$
\citep{2010A&A...515A..25D,2011MNRAS.415..235H,2014ApJ...790...39S},
larger by up to an order of magnitude than the gamma-ray luminosity
shown in the radio pulsar state. The brighter gamma-ray output
observed from MSPs in the intermediate disk state has been interpreted
in terms of an intra-binary shock close to the pulsar
\citep{2014ApJ...790...39S,2014MNRAS.444.1783C}, inverse-Compton
scattering of UV disk photons by the pulsar wind
\citep{2014ApJ...785..131T,2014ApJ...797..111L}, and synchrotron
self-Compton emission from the inner disk boundary around a propelling
NS \citep{2014MNRAS.438.2105P}. On the other hand, the lower
luminosity observed from the proposed counterpart of {\src} is similar
to that usually observed from MSPs in the rotation-powered state, and
indicates that this is the most likely state in which {\src} lies
during X-ray quiescence.

A detection of gamma-ray pulsations from \src\ would imply
rotational-powered activity in quiescence mode, whereas for
\src\ pulsed emission due to accretion-power mechanisms was detected
during the bursting accretion phase. If confirmed, SAX J1808.4$-$3658
will add to IGR J18245$-$2452 \citep{2013Natur.501..517P}, PSR
J1023+0038 \citep{2014arXiv1412.1306A} and XSS J12270$-$4859
\citep{2014arXiv1412.4252P} as a source showing evidence of a
transition to a rotation-powered radio pulsar state in X-ray
quiescence, whilst it is observed as an accreting pulsar when it has a
disk. That would also emphasise the potential of the gamma-ray regime
to investigate these systems, avoiding observational biases suffered
in other bands, such as large absorption or narrow radio beams.

\section{Conclusions}  
 
In a search for a gamma-ray counterpart for \src\ we discovered a weak gamma-ray source when analyzing almost six years of data obtained with
the LAT experiment. The position of the source is compatible within
3.2\arcmin\ with the location of \src.  The LAT source exhibits an
energy flux of ($2.1\pm0.5)\times10^{-12}$ erg cm$^{-2}$
s$^{-1}$ (in the 0.6 GeV to 10 GeV energy range) and a point morphology.

The positional alignment between the gamma-ray source and \src\ and
the lack of gamma-ray accelerators other than the AMSP suggest an
association between the two. However the uncertainties in the position and rotational ephemeris of \src\ prevent a firm
identification through phase variability. The uncertainty on the
  spin and spin frequency derivative will be improved by X-ray studies
  of the pulsations of the source during its future X-ray
  outbursts. On the other hand, the positional error is dominated by
  the $0.15\arcsec$ uncertainties on the 2MASS catalogue
  \citep{2006AJ....131.1163S} used to register the image of the
  optical counterpart of SAX J1808.4$-$3658
  \citep{2008ApJ...675.1468H}, and will hopefully be improved by future
  missions devoted to astrometry.

\section*{Acknowledgements}
Work done in the framework of the grants
  AYA2012-39303, SGR2009--811 and SGR2012--1073. N. R. acknowledges support from an NWO Vidi Award, and NewCOMPSTAR COST Action MP1304. A. R. acknowledges
  Sardinia Regional Government for financial support (P.O.R. Sardegna
  F.S.E. Operational Programme of the Autonomous Region of Sardinia,
  European Social Fund 2007-2013 - Axis IV Human Resources, Objective
  I.3, Line of Activity I.3.1). J.L. and D.F.T. acknowledge support from the National
  Natural Science Foundation of China via
  NSFC-11473027. D.F.T. further acknowledges the Chinese Academy of
  Sciences visiting professorship program 2013T2J0007.  

The \textit{Fermi} LAT Collaboration acknowledges generous ongoing support
from a number of agencies and institutes that have supported both the
development and the operation of the LAT as well as scientific data analysis.
These include the National Aeronautics and Space Administration and the
Department of Energy in the United States, the Commissariat \`a l'Energie
Atomique
and the Centre National de la Recherche Scientifique / Institut National
de Physique
Nucl\'eaire et de Physique des Particules in France, the Agenzia Spaziale
Italiana
and the Istituto Nazionale di Fisica Nucleare in Italy, the Ministry of
Education,
Culture, Sports, Science and Technology (MEXT), High Energy Accelerator
Research
Organization (KEK) and Japan Aerospace Exploration Agency (JAXA) in Japan,
and
the K.~A.~Wallenberg Foundation, the Swedish Research Council and the
Swedish National Space Board in Sweden.

Additional support for science analysis during the operations phase is
gratefully acknowledged from the Istituto Nazionale di Astrofisica in
Italy and the Centre National d'\'Etudes Spatiales in France.

\bibliography{ms_j1808}

\end{document}